\documentclass[secnumarabic,amssymb,nobibnotes,aps,prd,showpacs]{revtex4}%
\usepackage{graphicx}
\usepackage{dcolumn}
\usepackage{bm}
\usepackage{amsmath}%
\usepackage{amsfonts}%
\usepackage{amssymb}
\usepackage{color}
\usepackage{soul}

\begin{document}


\title{Ground level observations of relativistic solar particles on Oct 29th, 2015: Is it a new GLE on the current solar cycle? }

\author{C. R. A. Augusto, C. E. Navia, M. N. de Oliveira}
\affiliation{Instituto de F\'{\i}sica, Universidade Federal Fluminense, 24210-346,
Niter\'{o}i, RJ, Brazil} 

\author{A. A. Nepomuceno}
\affiliation{Departamento de Ci\^{e}ncias da Natureza, Universidade Federal Fluminense, 28890-000, Rio das Ostras, RJ, Brazil}

\author{A. C. Fauth}
\affiliation{Instituto de F\'{i}sica Gleb Wathagin, Universidade Estadual de Campinas, Campinas, SP Brazil}



\date{\today}
\begin{abstract}
On Oct. 29th, 2015, the Earth crossed  through a fold in the heliospheric current sheet. This is called a ``solar sector boundary crossing''.
Under this circumstances, a large coronal mass ejection (CME) occurred at 2:24 UT, behind the west limb on the sun. Therefore, the boundary crossing occurred when in the blast's nearby environment was filled with energetic particles accelerated by the CME shock waves, spacecraft measurements (ACE and GOES) have shown that in such a case, protons with energies at least up to 30 MeV were stored within the range of the sector boundary. Thus, a fraction of the solar energetic particles (SEP) from CME, reached Earth around 03:00 UT in the aftermath of the solar blast, reaching the condition of an S1 (minor) radiation storm level. 
The effect at ground level was a small increase in the counting rate in some ground based detectors, such as the South Pole Neutron Monitor (NM) and a sharp peak observed in the counting rate in the New-Tupi detector in Rio de Janeiro, Brazil and Thule NM. The event is being classified as a new GLE (Ground Level Enhancement) in the current solar cycle, as the GLE 73. However, in all cases, the counting rate increase is smaller or near than 2\%. 
The Earth crossed through a fold in the heliospheric current sheet also caused a  geomagnetic disturbance, below the minor geomagnetic storm threshold, observed  in the ACE spacecraft and a small decrease in the counting rates of some ground level detectors, such as the New-Tupi detector and Thuly NM.
Details of these observations are reported.

\end{abstract}


\maketitle

\section{Introduction}

 Cosmic rays ground-based detectors have been increasingly registering ground-level enhancements (GLEs) in the MeV and GeV ranges, leading to a very active research area.
Ground based instruments detects a large variety of secondary particles  produced by external 
primary protons (ions) penetrating the Earth’s atmosphere \cite{simpson00}. In most cases, GLEs are linked with the solar activity \cite{oh12}, such as the CMEs and their associated radiation storms \cite{firoz10}, that is, solar energetic particles. 

The register of GLEs as a signature of the solar activity has been very low in the current solar cycle 24. In the previous cycle 23, sixteen GLEs were observed \cite{gopal12}, while only a single event 
was observed, at least so far, in the current cycle: the GLE on May 17, 2012, linked with a M5.1-class flare and associated to a
CME and radiation storms. There is, however, one additional event detected with low confidence level by South Pole Neutron Monitor (NM) 
claimed as GLE and classified as GLE 72 \cite{thakur14}.

The detection of a GLE requires some favourable conditions, such as a geo-effective condition of the solar active region from where the solar flare and its associated CME originate. In most cases only solar active regions, located around the middle west sector of the solar disc, has triggered GLEs.
In addition, the detection  of a GLE depends of the detector location. For instance, 
the GLEs observed in the solar cycle 23 were observed by detectors located at regions with high latitudes, 
where the geomagnetic rigidity cutoff is around 1GV or smaller \cite{andriopoulou11}.

In this paper we report results from a new ground based (5 m above sea level) detector, 
denominated as the New-Tupi telescopes. They are located at Niteroi city, Brazil ($22.9^0$W, $43.2^0$S), 
within the South Atlantic Anomaly (SAA), a region characterised by an anomalous low strength geomagnetic field when 
compared to the geomagnetic dipole field. The main goal of the New-Tupi telescopes is the characterization of space 
weather effects at ground level.

We have identified particles (muons) in excess on New-Tupi detector (a sharp peak) on Oct 29, 2015, that correlates with the onset of a radiation storm, as well as with 
ground-based observations made by detectors located at high latitudes, such as the South Pole NM, that is one of the eight NMs successfully operated by the Bartol Research Institute
\cite{bartol1}. The South Pole NM is strategically located at South pole, at 90 degrees of latitude, where the geomagnetic rigidity cutoff is 0.1 GeV, the lowest in the world. The signal was also observed in the IceTop, an air shower detector located on the top of the IceCube neutrino observatory at South Pole. Despite to the confidence level, in all cases not bigger  than 2\%, it is claimed that the event could  be a new GLE 
on the current solar cycle 24 and classified as the GLE 73 \cite{bartol2}. 

The event on Oct 29, 2015 was a unusual event because there  is no register of X-ray emission due to a solar blast. However, a partial-halo CME was observed in coronagraph imagery Lasco C2 on board of the SOHO spacecraft \cite{robb04}. That happened behind the west limb on the sun, that is, without an apparent magnetic connection with the Earth, but even so, 
the CME  has triggered a radiation storms on Earth on October 29, 2015.

As the solar sector boundary crossing occurred when in the blast's nearby environment on Oct 29, 2015 was filled with energetic particles accelerated by the CME shock waves, we argue that the SEP observed at GOES and also  observed at ground level, 
was guided from the sun to the Earth through the fold of 
current sheet \cite{wilcox65}

​We also reported the effects  due to the Earth crossing the boundary region causing short-lived geomagnetic disturbances in the IMF \cite{svestka68} observed  in the ACE spacecraft and a small decrease in the counting rates of some ground level detectors, such as the New-Tupi, on Oct 29, 2015. In addition, a brief report on the local geomagnetic field 
effect on the sensitivity of a ground-based detector is presented.

This paper is organized as follows. In Section 2, we give a brief description of the local geomagnetic effect on the sensitivity of ground-based detectors. The aim is  to show why the New-Tupi detector has a sensitivity close to the sensitivity of the detectors located at high latitude. Section 3 is devoted to report the observations, including in subsection 3.1, the WSA-Enlil prediction model of the heliosphere, showing of solar wind density 
on Oct 29, 2015, and the Earth crossing  through a fold in the heliospheric current sheet.
Spacecraft observations on CME and their associated shock waves, and their 
effects such as the radiation storm is presented in subsection 3.2
The ground level observations, specially the New-Tupi observations and their 
 correlation with the observations reported by space-borne detectors and other ground based experiments is presented in subsection 3.3.
 
In section 4 we present the effects on the geomagnetic field due to the passage of the Earth by the      boundary region of the current sheet. A small geomagnetic disturbance, below the minor geomagnetic storm threshold, observed in space-borne detectors and ground level detectors and the section 5 present our conclusions.
The paper has a appendix where a brief description of 
the New-Tupi detector is presented.

\section{Local geomagnetic field effect on the sensitivity of a detector}



Secondary particle such as pions are produced in nuclear interactions of solar particles (above GeV energies) in the upper atmosphere.
Thus, a particle such as a electron or a muon produced by pion decay is characterized by a 
momentum $p=(p_{\parallel}^2+p_{\perp}^2)^{1/2}$, where $p_{\perp}$ is the transverse momentum and $p_{\parallel}$ 
is the longitudinal momentum.
Particles with a vertical momentum $p$, produced at the altitude z, have a lateral spread at observation level as
\begin{equation}
r=\left( \frac{p_{\bot}}{p_{\parallel}}\right) z.
\end{equation}

The transverse momentum is the main cause of the angular divergence of secondary particles. The lateral spread by this effect is large and increases as the particle's momentum decreases.
In addition, charged particles such as electrons and muons traversing a finite thickness of matter suffer repeated elastic Coulomb scattering.
The cumulative effect of these small angle scattering is a net deflection from the original particle direction.
The mean-squared angle of multiple scattering of a muon with energy $E_{\mu}$ can be approximated \cite{olbert54} as

\begin{equation}
<\theta^2>^{1/2} \approx \sqrt{5.8\;MeV/E_{\mu}},
\end{equation}
this is about $4.4$ degrees for a muon with $E_{\mu}=1.0$ GeV.

However, there is an additional source for the lateral dispersion of secondary charged particles  that is neglected in most cases. This additional lateral effect is due to the Earth's magnetic field \cite{reyes05,augusto15}.
Let us consider a bundle of protons ($E_p>1$ GeV) incident vertically  at the top of the atmosphere.
The charged particles produced in the atmosphere tend to travel in the downward direction. 
As the charged particle travels, its initial directions is  shifted by a horizontal distance $\Delta r$, in the direction  perpendicular to the Earth's magnetic field
 $B_{\bot}$
\begin{equation}
\Delta r \sim z^2/R = z^2ceB_{\bot}/p, 
\end{equation}
where $z$ is the height of the atmosphere where the charged particle is generated, 
and $R$ is the radius of curvature of an (initially) vertical charged particle (with charge $e$) travelling
downward in the atmosphere with momentum $p$.

Thus, the deflection by the geomagnetic field of charged particles during their propagation in the atmosphere is caused by the component of the geomagnetic field perpendicular to the particle trajectory. This effect results in a decrease 
in the number of collected particles and therefore the detector sensitivity. 
The sensitivity of a detector due to geomagnetic field effect is inversely proportional 
to the transverse component of the geomagnetic field to the particle trajectory.

From the horizontal and the vertical components of the geomagnetic field, ($B_h, B_z$), in a given location in which the detector operates, it is possible to obtain the transverse magnetic component for the particle propagation trajectory, for any direction, specified by the zenith $\theta$
and azimuth $\phi$ angles: 
\begin{equation}
B_{\bot}^2=B_h^2+B_z^2-(B_h\sin \theta \cos \phi + B_z \cos \theta)^2,
\end{equation}

We have analysed graphically the transverse component, $B_{\bot}$ described above for four different sites, where there are particle detectors, with different geomagnetic rigidity cutoff, including the New-Tupi location, within the SAA region. The geomagnetic parameters of these sites  are summarized in Table 1.

\begin{table}[!h]
\centering
\caption{Geomagnetic parameters of the four places analysed here. \cite{noaa}.
}
\label{my-label}
\begin{tabular}{cccccc}
\hline
\hline
Site & \begin{tabular}[c]{@{}c@{}}(Latitude, Longitude)\\ (degree)\end{tabular} & \begin{tabular}[c]{@{}c@{}}B(horizontal)\\ (mT)\end{tabular} & \begin{tabular}[c]{@{}c@{}}B(vertical)\\ (mT)\end{tabular} & \begin{tabular}[c]{@{}c@{}}B(total)\\ (mT)\end{tabular} & \begin{tabular}[c]{@{}c@{}}Stormer vertical \\ rigidity cutoff (GV)\end{tabular} \\ 
\hline
\hline
South Pole   & 90.0S                                         & 16.74                                                      & -52.26                                                                      & 54.9                                                  & 0.1                                                  \\ 
Thule     & 76.5N, 68.7W                                        & 4.52                                                    & 56.16                                                                      & 56.34                                                  & 0.3                                                  \\ 
Athens   & 37.9N, 23.7E                                      & 26.68                                                   & 37.36                                                                      & 45.91                                                  & 8.5                                                 \\ 
New-Tupi   & 22.9S, 43.1W                                      & 17.89                                                   & -14.81                                                                      & 23.22                                                  & 9.2                                                 \\ \hline
\end{tabular}
\end{table}

From Table 1, we can see that among the four places, the New-Tupi location has the smaller total geomagnetic field, 
(B(total)). However, it is also the place with the higher Stormer rigidity cutoff (9.2 GV),
that is, in the dipole approximation of the geomagnetic field 
and that does not include the magnetic anomaly (SAA).

If we consider only particles with vertical incidence (see third column in Table 1), 
a detector located in Thule would have the highest sensitivity, 
because is the place with the smallest magnetic horizontal component. 
In this case, the sensitivity of a detector located at New-Tupi site would be close to 
the sensitivity of the detector located at South Pole. Fig. 1 show the perpendicular 
magnetic component to the particle's trajectory to other directions, that is, as a function of 
the pointing direction $\theta$ and $\phi$.

\begin{figure}
\vspace*{+0.0cm}
\hspace*{0.0cm}
\centering
\includegraphics[width=11.0cm]{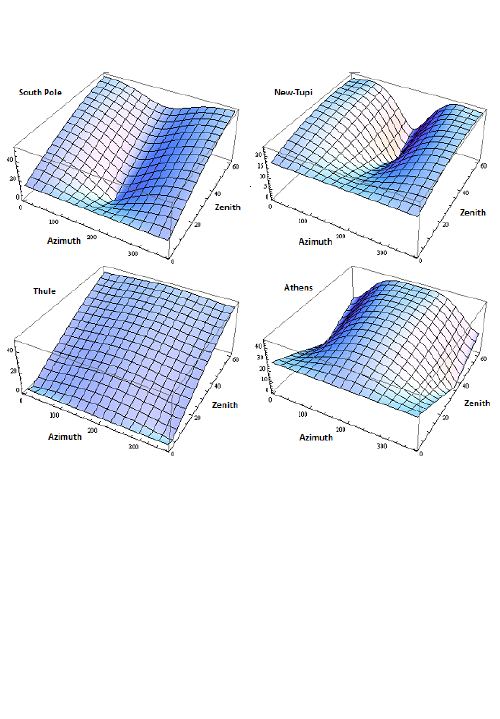}
\vspace*{-5.0cm}
\caption{Perpendicular 
magnetic component,  $B_{\bot}$, to the particle's trajectory  as a function of 
the pointing direction $\theta$ (zenith) and $\phi$ (azimuth), for four places: South Pole, New-Tupi, Thule and Athens, respectively.
}
\label{fig1}
\end{figure}

\begin{figure}
\vspace*{-1.0cm}
\hspace*{0.0cm}
\centering
\includegraphics[width=8.0cm]{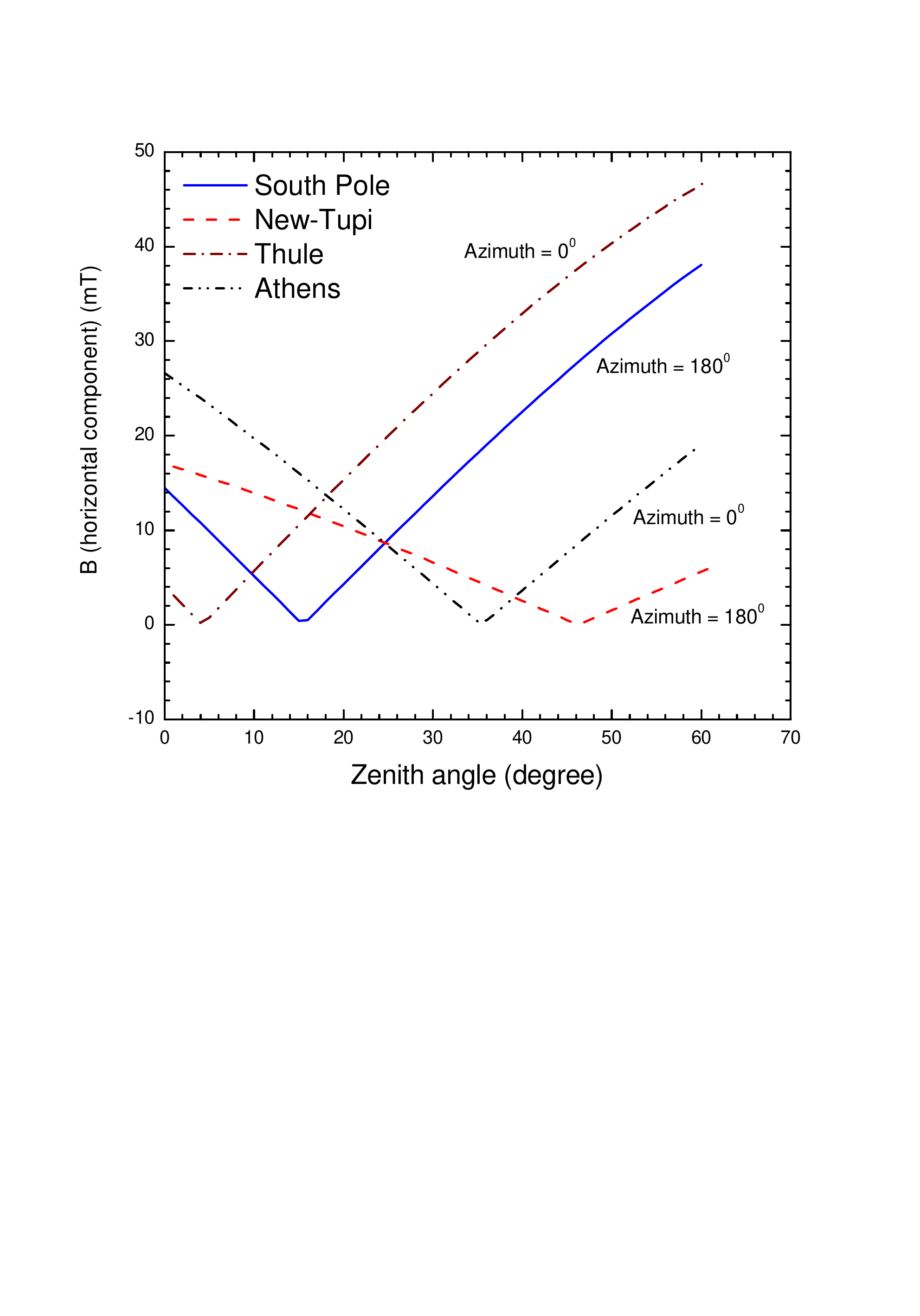}
\vspace*{-4.5cm}
\caption{Perpendicular 
magnetic component,  $B_{\bot}$, to the particle's trajectory  as a function of 
the zenith angle, $\theta$, and fixed azimuth angle: $\phi=0^0$ for the two places (Thule and Athens) located  in the North Hemisphere, and $\phi=180^0$ for the two places (South Pole and New-Tupi) located in the South Hemisphere,  respectively.
}
\label{fig2}
\end{figure} 

Clearly, a funnel effect can be seen. 
For detectors installed in the southern hemisphere, the maximum sensitivity is achieved for particles coming 
from the south ($ \phi = 180^0$), and for the detectors installed in the northern hemisphere, the maximum sensitivity is 
achieved for particles coming from the north ($ \phi =0^0 $ ). However, this maximum sensitivity is achieved for 
different zenith angles, as best shown in Fig. 2.

In all cases, $B_{\bot}$ decrease as the zenith angle increases, up to the point of their maximum sensitivity, $B_{\bot}=0$. From there, 
 the $B_{\bot}$ increases as the zenith angle increases. 
We can also see that  of increase or decrease rate of the sensitivity with the 
zenith angle is the same in the South Pole, Thule and Athens. The exception is the New-Tupi location, where  the increase (decrease) of these rates are less accentuated, that is, an anomalous behaviour.

It is possible to see that for a vertical incidence or near to this, the location of Thule presents greater sensitivity, but for zenith angles above $ 5^0$, 
the sensitivity rapidly decreases as the zenith angle increases.

 The sensitivity in the South Pole increases rapidly with increasing zenith angle,  up to 15 degrees, where it reached the maximum sensitivity and from there the sensitivity decreases 
with increasing zenith angle.

A similar behavior can be seen for the case of Tupi location. However, in this case, as already comment, 
the sensitivity increases more slowly with increasing zenith angle, reaching the maximum sensitivity only for 
zenith angle of 47 degrees. We point out that for a vertical incidence, the sensitivity in New-Tupi site is 
very close to the sensitivity at the South Pole.

Among the four places, and for vertical incidence or around it ($\theta < 24^0$), the Athens's  location has the bigger $B_{\bot}$ (the smaller sensitivity). 
In the region $\theta < 24^0$, the sensitivity at Athens is  smaller than the sensitivity at New-Tupi, despite  the vertical rigidity cutoff be smaller at Athens.   However, this behavior at Athens is useful to the study of solar transient events with a relatively high intensity. 
This local geomagnetic field effect is taken into account during shower development in the atmosphere, in Monte Carlo calculations such as the FLUKA. 

On the other hand, the point of first interaction of primary particles is also modulated by the geomagnetic field, this introduces a geomagnetic cutoff, in this case the horizontal component,
$B_{\bot}$, to particle trajectory, plays also a important role. This effect is also included in Monte Carlo calculations,such as FLUKA.

\section{Observation}

\subsection{The Earth boundary crossing on Oct 29, 2015}

Heliospheric current sheet (HCS) is a thin transition zone where the polarity of the sun's magnetic field changes from plus (north) to minus (south). 
\cite{wilcox65,israelevich01,foullon09}.
In other words, the  heliospheric current sheet is  a thin region of space with no magnetic 
polarity and a large electrical current, dividing the inward polarity and outward polarity regions \cite{mursula03}.

The current sheet is important because charged particles, such as solar particles and low energy 
cosmic rays, tend to be guided by its folds. This is the reason why the particle density in the fold is always 
higher than the density out of it.  For instance, at 1 AU the current sheet can produce auroras at polar 
regions such as the produced by CMEs and coronal holes that increase the solar wind.

\begin{figure}
\vspace*{-1.0cm}
\hspace*{0.0cm}
\centering
\includegraphics[width=6.0cm]{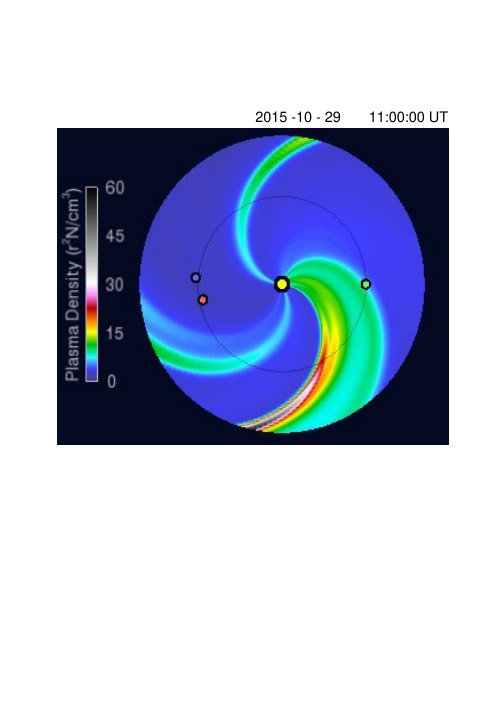}
\vspace*{-3.0cm}
\caption{ The WSA-Enlil large-scale, physics-based prediction model of the heliosphere, showing of solar wind density 
on Oct 29, 2015, and the Earth crossing  through a fold in the heliospheric current sheet. The Sun is represented as a yellow dot, the Earth by a green dot, and the STEREO spacecraft by the red and blue dots.   Credit: SWPC.
}
\label{fig3}
\end{figure} 

The heliospheric current sheet passage near Earth is a relatively well predicted recurring phenomenon. A HCS past the Earth every 27 days (rotation rate of the sun). But it can has different structure after one rotation and  substructures caused by out flowing transients. In addition, the main orientation can be highly distorted. The geomagnetic storms caused by these recurring features are relatively weak, in most cases below the minor geomagnetic storm threshold.

Fig. 3 shows the forecast of conditions in the solar wind density as predicted by the WSA-Enlil model on Oct 29, 2015. 
The picture also shows the Earth crossing through a fold in the heliospheric current sheet. 
Such condition has allowed a special magnetic connection between the Sun and the Earth.
This is evident by the fact that energetic particles from a wide CME that happened 
in the opposite side on the sun, but near the western limb, managed to reach the Earth.
The circular plots are a view from above the North Pole of the Sun. The Sun is the yellow dot in the center and the 
Earth is the green dot on the right.  The locations of the two STEREO satellites 
are also shown. Clearly, the figure shows the conditions  described above.

\subsection{The CME and radiation storm on Oct 29, 2015}

On Oct 29, 2015, the Earth facing solar activity remained low, without chances of a noteworthy 
solar flare, as shown in the upper left-hand corner of Figure 2, where the  GOES X-ray flux is shown. There is no sign of a solar flare that could trigger a CME. However, at 2:24 UT a
partial-halo CME was observed in the coronagraph imagery Lasco C2 instrument aboard the SOHO spacecraft.
The CME emission was in the south-west sector of the Sun, as show in the upper right-hand corner of Figure 2. 
Additional information provided by Solar Dynamics Observatory (SDO) shows 
that a noteworthy eruption occurred behind the west limb on the far side of the sun. 
On the other hand, the angular distribution of the principal angle (counter-clockwise from North) 
of the shock waves triggered by CME, provided by Cactus catalog, has shown that a fraction of 
them are located around 270 degrees, that is, on the ecliptic plane, as show in the lower right-hand corner of figure 2. 
This means that particles were accelerated by these shock waves and injected around the ecliptic plane.

Indeed, a proton flux increase was observed on Oct 29, 2015, around 3:00 UT on the ACE-SIS spacecraft (at Lagrange point L1), as well as, on the geostationary GOES satellite, reaching the condition of S1 level, of radiation storm in the NOAA storm scale (proton flux above 10 particles$/cm^2 s\;sr$, with energies above 10 MeV 
at 1 AU), as show in the lower left-hand corner of figure 2. 
This radiation  storm was the result of acceleration of particles by shock waves triggered by the CME 
that occurred from just beyond the SW limb of the Sun  around 02:19 UT.

\begin{figure}
\vspace*{-1.0cm}
\hspace*{0.0cm}
\centering
\includegraphics[width=13.0cm]{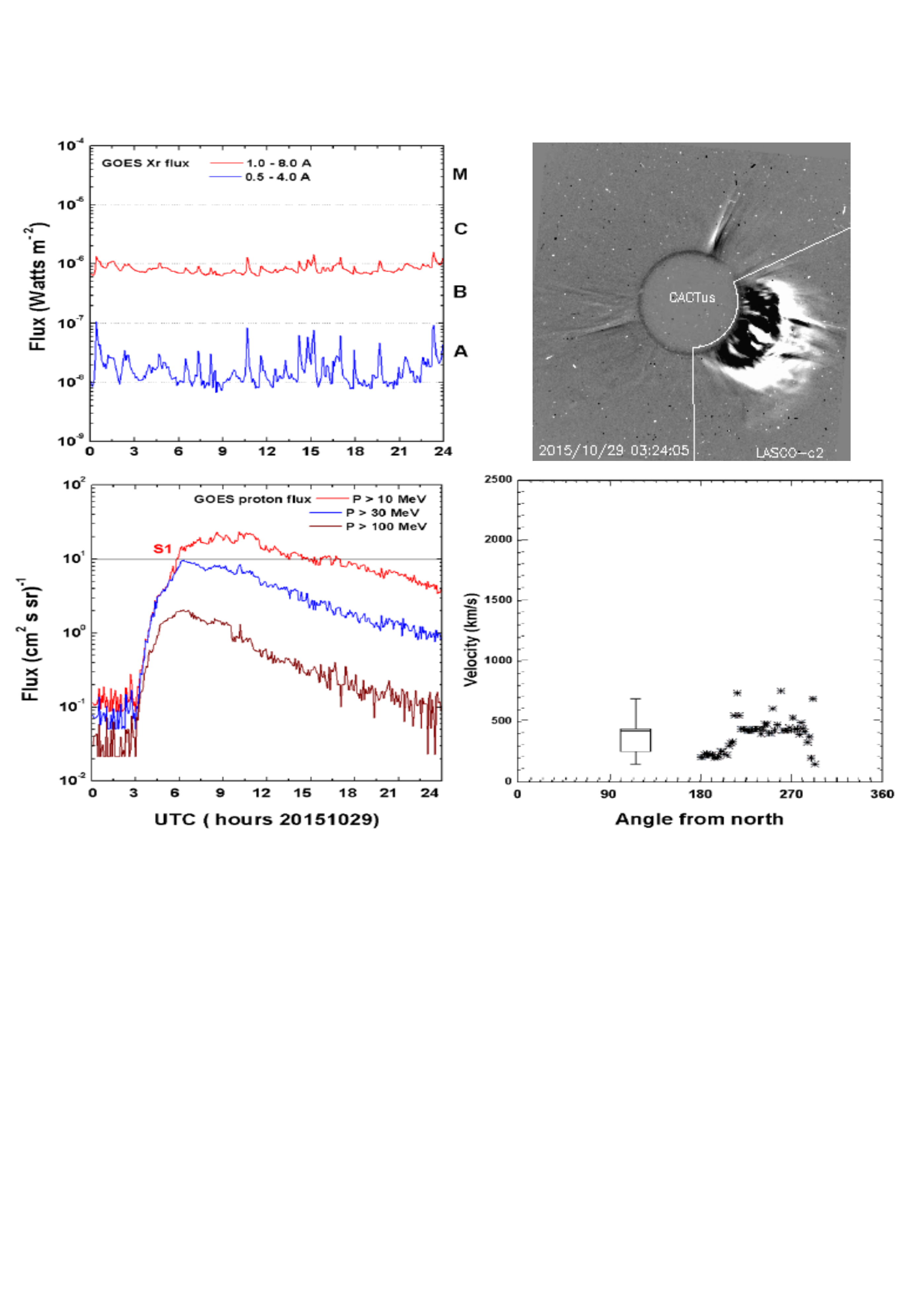}
\vspace*{-7.0cm}
\caption{Upper left corner: The GOES X-ray flux on Oct 29th, 2015 in two wavelength. Upper right corner: The Lasco C2 observation of the CME 0088 on Oct 29th, 2015 at 03:24:05 UT. Lower left corner: The GOES proton Flux  on Oct 29th, 2015 in three energy bands. Lower right corner: Correlation between the velocity of CME shock waves and the principal angle, counterclockwise from North, linked with the CME 0088.
}
\label{fig4}
\end{figure} 

\subsection{Ground level observations}

In  temporal coincidence with the onset of the radiation storm observed on the GOES satellite, 
there was an increase in the counting rate in the South Pole NM. The
South Pole NM has a strategic location, at latitude $90^0$ South, and it is the detector with the
lowest geomagnetic rigidity cutoff (0.1 MV) on Earth.  The particle enhancement at 
South Pole NM means the presence of particles around GeV energies region, probably in the tail of the energy spectrum. 
However, the increase was only up to 1.5\%, relative to the background, in the 10 minutes binnig count rate, 
and the duration of the enhancement was less than 4 hours, as shown in Fig. 5 (bottom panel).

A search in other detectors at ground level was made and a similar signal 
was found in New-Tupi detector, working in scaler mode, as is shown in Fig. 5 (central panel). 
The New-Tupi detector is located in a place where the Stormer geomagnetic rigidity cutoff is 9.2 GV. However, the  site is within of the SAA central region, 
and as  shown in section 2, this region has a high sensitivity to detect charged particles.

The confidence level at New-Tupi is above 2\%, but it is only a sharp peak, a behaviour
 that was also found at Thule NM, a detector located at high latitude on the North Hemisphere (see Table 1). That is, 
Thule NM has also a sharp peak with an increase of up to 2\% and temporally coincident with the New-Tupi peak. 
Fig. 6 summarizes the situation, where the time profiles 5 minutes binning of the New-Tupi, SOPO, 
Thule and McMurdo data are shown. Table 1 presents their coordinates and magnetic parameters.

We would like to point out that in all cases the particle enhancement was 
below that required to trigger a GLE alert. Indeed, two GLE alerts
have been developed, one of them  uses the eight Bartol NMs stations
 \cite{kuwabara06}. 
The signal in three of them (South Pole, Thule and MacMurdo) is shown in Fig. 6.  The other one is the GLE ALERT System
developed by the National and Kapodistrian University of Athens-Cosmic Ray Group \cite{mavro12}, 
which use the ESA NM stations (Souvatzoglou et al. 2009). 
Both systems use the same scale. On Oct 29, 2015, no alert was reported, 
since it is necessary that three or more stations observe an increase equal to or greater than 4\%.

\begin{figure}
\vspace*{-0.0cm}
\hspace*{0.0cm}
\centering
\includegraphics[width=10.0cm]{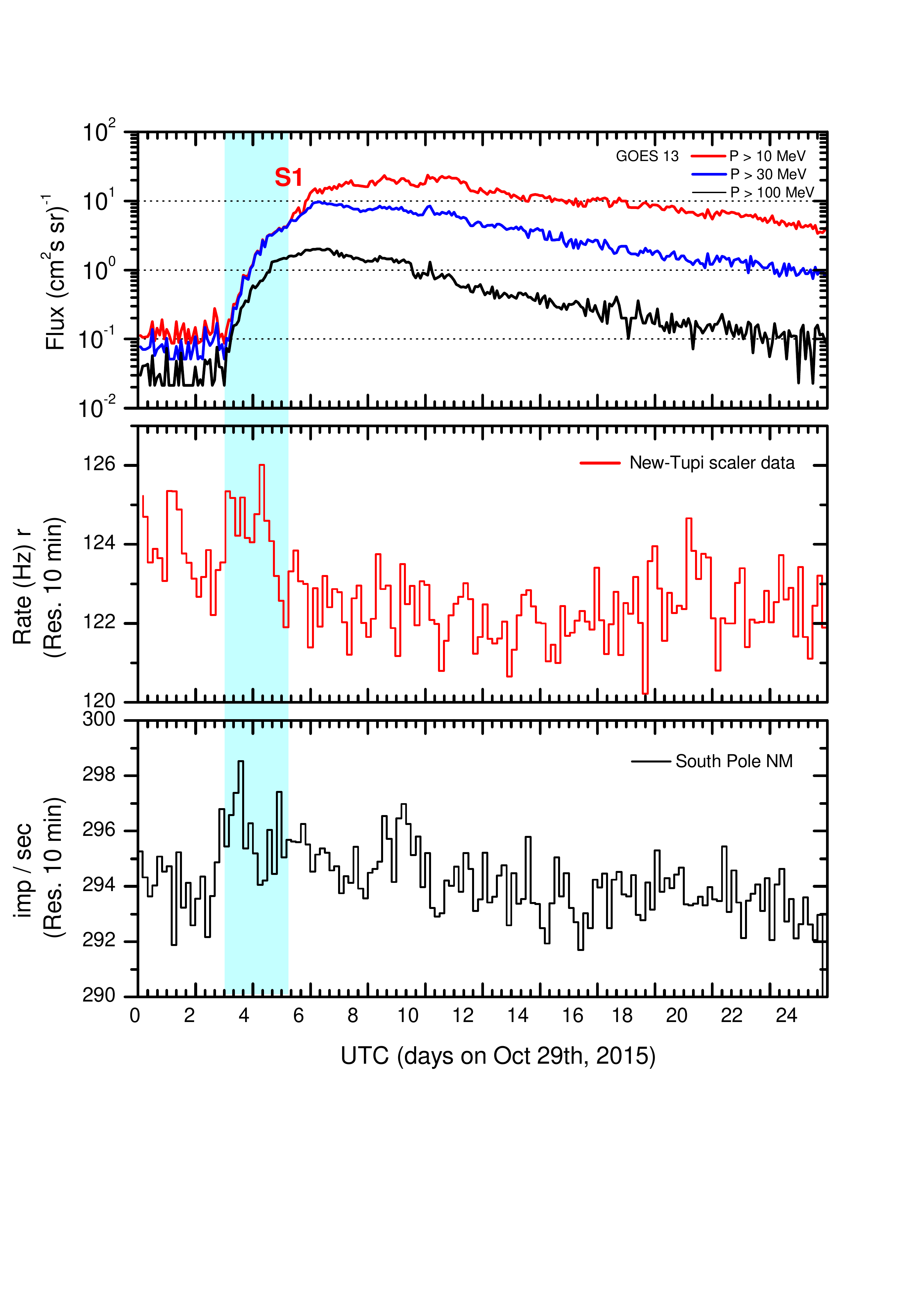}
\vspace*{-2.0cm}
\caption{The Time profiles on Oct 29, 2015. Top panel: The GOES 13 Proton flux in three energy bands.
Central panel: The muon counting rate (escaler mode) in the Tupi-New detector, 10 minutes binning. 
Bottom panel: The particle counting rate in the South Pole NM, 10 minutes binning.}
\label{fig5}
\end{figure} 

\begin{figure}
\vspace*{-0.0cm}
\hspace*{0.0cm}
\centering
\includegraphics[width=10.0cm]{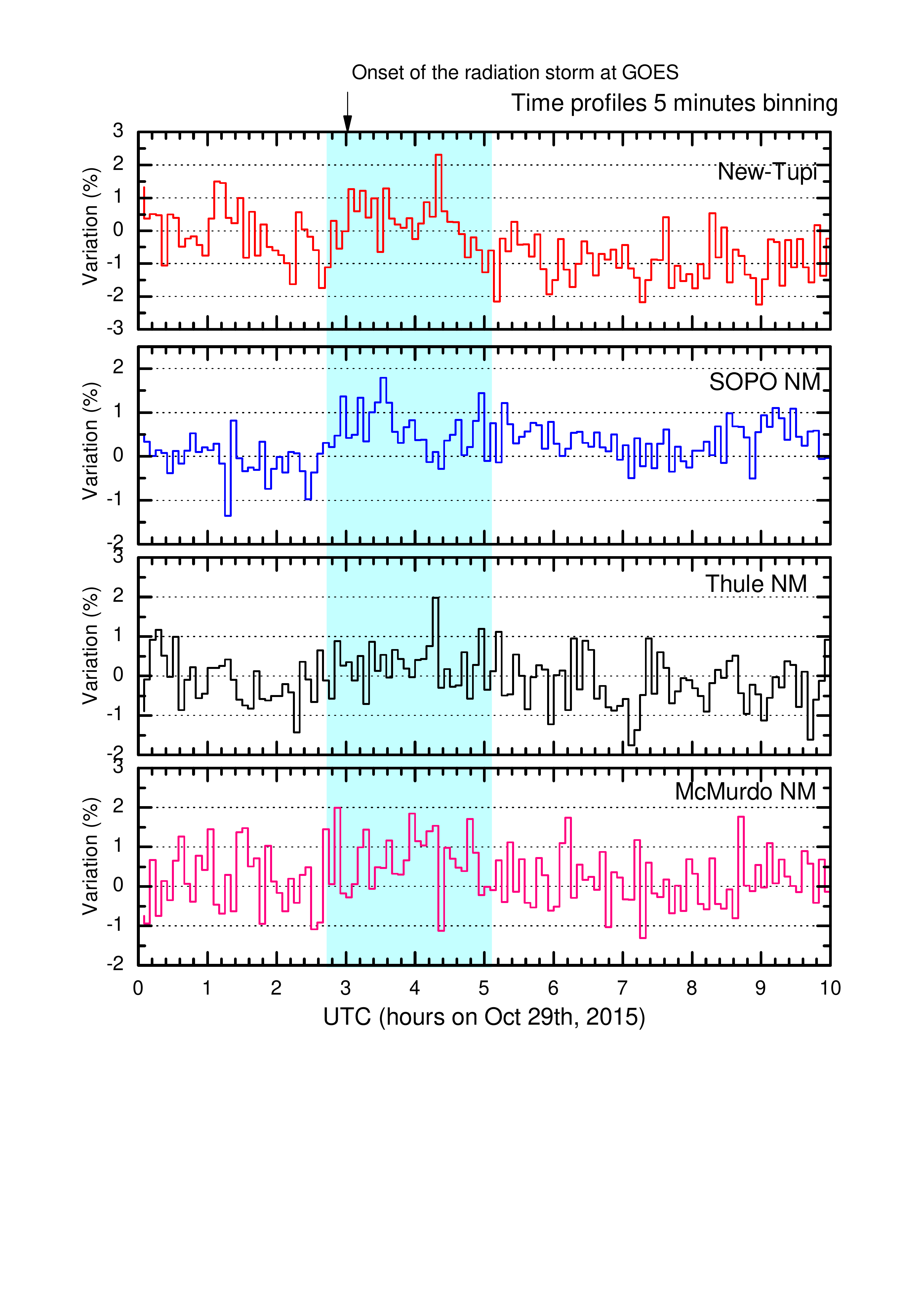}
\vspace*{-2.5cm}
\caption{The Time profiles expressed in percentage variation in the counting rates on Oct 29th, 2015. From top to bottom: the muon counting rate (escaler mode) in the Tupi-New detector, the particle counting rate in the South Pole NM (SOPO), Thule NM and McMurdo NM, respectively. In all cases 5 minutes binning.
}
\label{fig6}
\end{figure} 

\section{Geomagnetic disturbance due to  boundary crossing}

The passage of the Earth by the heliospheric current sheet is a relatively well predicted recurring phenomenon.
The Earth can pass a same HCS every 27 days,
that is the rotation rate of the sun, but the HCS can have different structures, as well as
substructures caused by out-flowing transients. In addition,
the main orientation can be highly distorted [e.g. Villante et al., 1979] \cite{villante79}.

The geomagnetic storms caused by these
recurring features are relatively weak, below the minor geomagnetic storm threshold and in most cases are long-lived.
The energy fluctuations of the ions, compressed within the field reversal regions, 
can be interpreted as Alfénic fluctuations and the main effect is a small variation in the $B_{z}$ component of the IMF
and can be weakly oriented southward, providing a magnetic reconnection with the Earth's magnetic field, oriented northward,  several times,  
but for short periods, with a 2 h average, opening cracks in the magnetosphere, promoting an injection of the solar wind in the upper atmosphere producing auroras in the polar regions.
\begin{figure}
\vspace*{-1.0cm}
\hspace*{0.0cm}
\centering
\includegraphics[width=12.0cm]{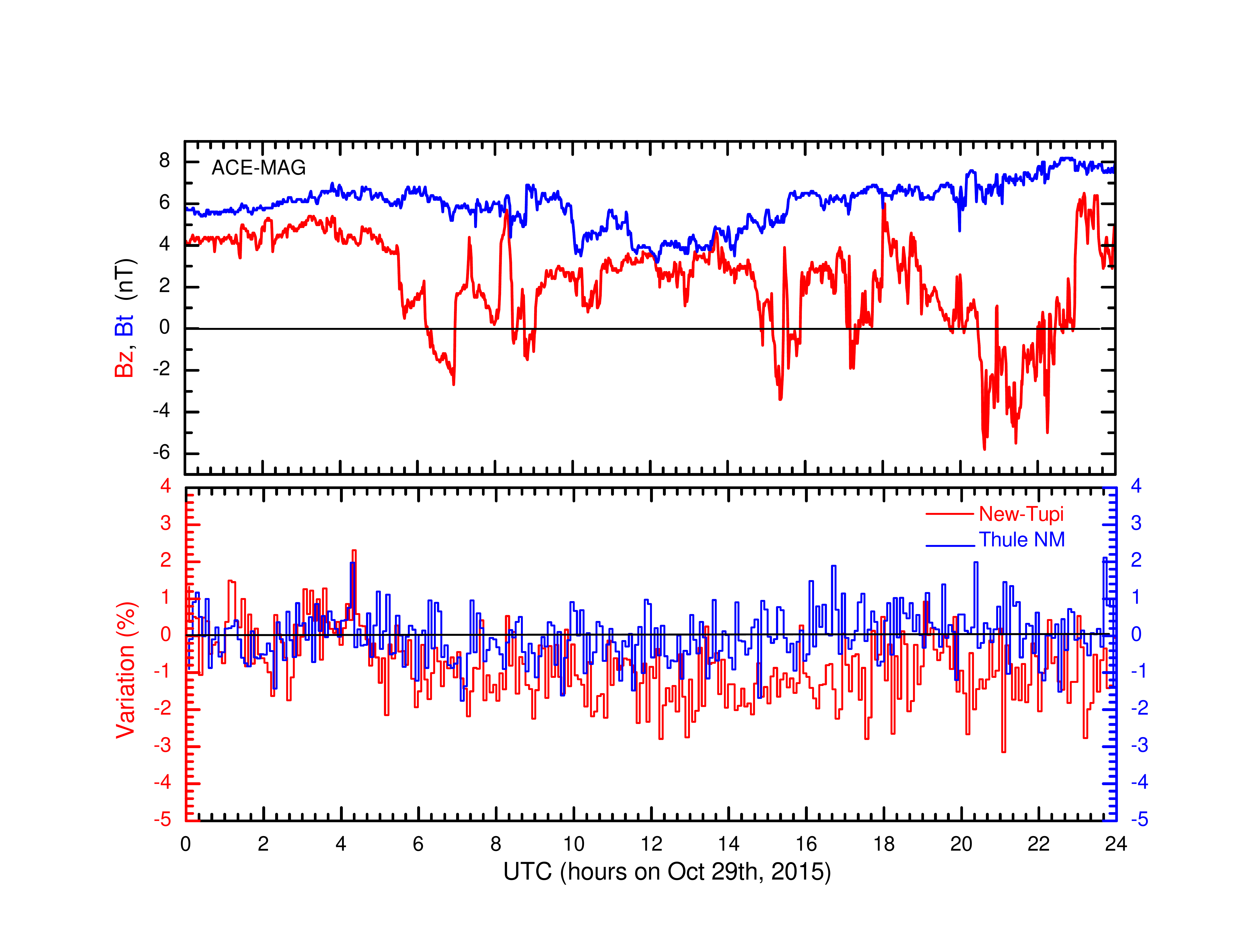}
\vspace*{-0.5cm}
\caption{ Time profiles on Oct 28th. Top panel: The Bt and Bz magnetic components of the local IMF, observed by the ACE-MAG instrument on-board the ACE spacecraft at Lagrange point L1.
Bottom panel: The muon counting rate (escaler mode) in the Tupi-New detector (red) and the particle counting rate in the Thule NM (blue), both 5 minutes binning. 
}
\label{fig7}
\end{figure} 

Fig. 7 (top panel) shows the results of measurement of the IMF carried out with ACE-MAGNETIC detector in the Lagrange point L1, on Oct 29th, 2015. As expected, about one hour after
of the arrival of the SEP in the Earth, that is, when the Earth began to cross the boundary 
sector of the heliospheric current sheet, the $B_{z}$ component also began to be oriented southward, 
and this coincides with a small fall in the muon counting rate in the New-Tupi detector,   and in the Thule NM, as shown in Fig. 7 (bottom panel), respectively). 
However, in all cases the confidence level is not more than
1.5 \%. This ``mini-Forbush decrease'' occurred immediately after the sharp peaks observed in both detectors 
and they represent the particle excess due to the arrival at Earth of SEP.

From  Fig. 7 we can see also that there are at least other four oscillations in the $B_{z}$ component on Oct 29, 2015. 
During this period the counting rate remains in the same level
after the fall, but with some small oscillations.

Thus, the geomagnetic effects caused by this heliospheric boundary crossing was relatively weak, 
below the minor geomagnetic storm threshold, without an alert
through of the geomagnetic indexes, such as the Kp and Dst.

\section{Conclusions}

The SEP fluxes derived from the data of ground-based detectors are model dependent and require knowledge of response function. Even so, only the ground-based 
installations can provide the sufficient geometric factor to measure the low 
intensities of the solar protons of highest energy. Thus, ground-based measurements 
are complementary to the measurements performed on spacecraft. To this end 
there are several world-wide networks of ground-based detectors, such as the
standard NMs, that are  useful instruments
with a huge geometric factor to observe temporal variations of galactic cosmic rays
 and SEP with energies above $\sim 1$ GeV. In particular, we would like to highlight the Bartol's NMs located at high latitudes (polar regions), consequently, they have the highest sensibilities.

Within the ESA Neutron Monitor Database (NMDB) we would like to highlight the Oulu NM with stable data since 1966. The Oulu NM, located at Finland,  has a geomagnetic rigidity cutoff of 0.8 GV \cite{usoskin}. In the Oulu NM web page, there is an useful GLE catalogue \cite{oulu}, with information since the GLE 15, occurred on July 07, 1966. The last record in the catalogue is the GLE 71 on May 17, 2012, the only GLE in the current solar cycle 24.

However, there are at least other two events that are being considered as GLEs. The first
was on January 6, 2014, catalogued as the second GLE in the solar cycle 24. This event was primarily observed by the South Pole NM (increase of $\sim 2.5\%$), whereas a few other neutron monitors recorded smaller increases \cite{thakur14}. Even so, the event is being classified as the GLE 72.

The second event that is being classified as the GLE 73 is the event on Oct 29, 2015, analysed here. As the previous GLEs, the event was observed only by some ground-level detectors located at high latitudes (polar regions), where the geomagnetic rigidity cutoff is lower. However, in this case, there is also signal at a ground-based detector at 
median latitude (the New-Tupi detector). We have shown (see section 2) 
that despite  the relative high rigidity cutoff (9.2 GV), the detector 
is located within the SAA region, where the local geomagnetic field is the smallest in the world and therefore it allows a detector has high sensitivity to charged particles.
 We have also shown (see section 3.1) that the event has occurred when the Earth crossed through a fold in the heliospheric current sheet and a fraction of the high energy particles triggered by the solar blast were directed to the Earth, through the fold of the heliospheric current sheet. Without this condition, probably would not exist the particle enhancements at spacecraft and ground level detectors.
 
 In most cases, SEP observations imply the existence of
coronal shocks extending at least $\sim 300^0$ \cite{cliver95} and IP shocks up to $\sim 180^0$ at 1 AU \cite{cane96}. For instance, in the solar cycle 23,  most (80\%) of the CMEs correlated with GLEs were full halo type, with three events non-halos, even so,
having widths in the range 167 to 212 degrees \cite{gopal12}. Whereas, in the event on Oct 29, 2015, the CME had only an angular width of 114 degrees (see the upper right corner in Fig. 4). In addition, as shown in section 3.3, the particle enhancement observed at ground level,
in all cases has a confidence level around 2\% or smaller than this. even so,
it is being considered as the GLE 73. 

If we used these criteria, we can claim that there are, at least,  other two events that can be classified as GLEs in the current solar cycle, with results already published. They are the event  reported as 
\textit{Signals at ground level of relativistic solar particles associated with a radiation storm on 2014 April 18} \cite{augusto15b}, and the event reported as \textit{Signals at ground level of relativistic solar particles associated to the "All Saints" filament eruption on 2014} \cite{augusto15c}. In both cases, the signals of several NMs are reported, including the South Pole NM and the Tupi telescope (the  predecessor of the New-Tupi detector). Also in both cases, the signals have  confidence levels above 3\%.

\appendix
\section{The New-Tupi detector} 
\label{App:AppendixA}

The aim of the New-Tupi detector is the study of the space weather effects at ground level, driven by diverse solar transient events.  The
New-Tupi telescope is located at Niteroi city, Rio de Janeiro
state ($22^0 53'00''S,\; 43^0 06'13'W$) and at 5 m a.s.l, that is, within 
the South Atlantic Anomaly (SAA) region, and it is the
region where the inner Van Allen radiation belt makes its closest approach
to the Earth’s surface. As a consequence of this behaviour, the SAA region is
characterized by an anomalously weak geomagnetic field strength (less than
28,000 nT) \cite{barton97}.

The New-Tupi detector has four detectors assembled on the basis of
plastic scintillator (150cm x 75cm x 5cm) within a trapezoidal truncated box,
and a Hamamatsu, model R877, with 10 stages, 127 millimeters of
diameter photomultiplier is assembled at the truncated part. 
The data acquisition system is made on the basis of an National Instrument DAQ
with eighth analogical channels, 1 MS/s, 12-bit, simultaneous sampling rate.

 Only signals above a threshold value, corresponding
to energy of about 100 MeV deposited by particles (mostly muons) that
reach the detector, are registered. The output raw data of each detector is
recorded at a rate of one Hz.

\begin{figure}
\vspace*{-4.0cm}
\hspace*{0.0cm}
\centering
\includegraphics[width=13.0cm]{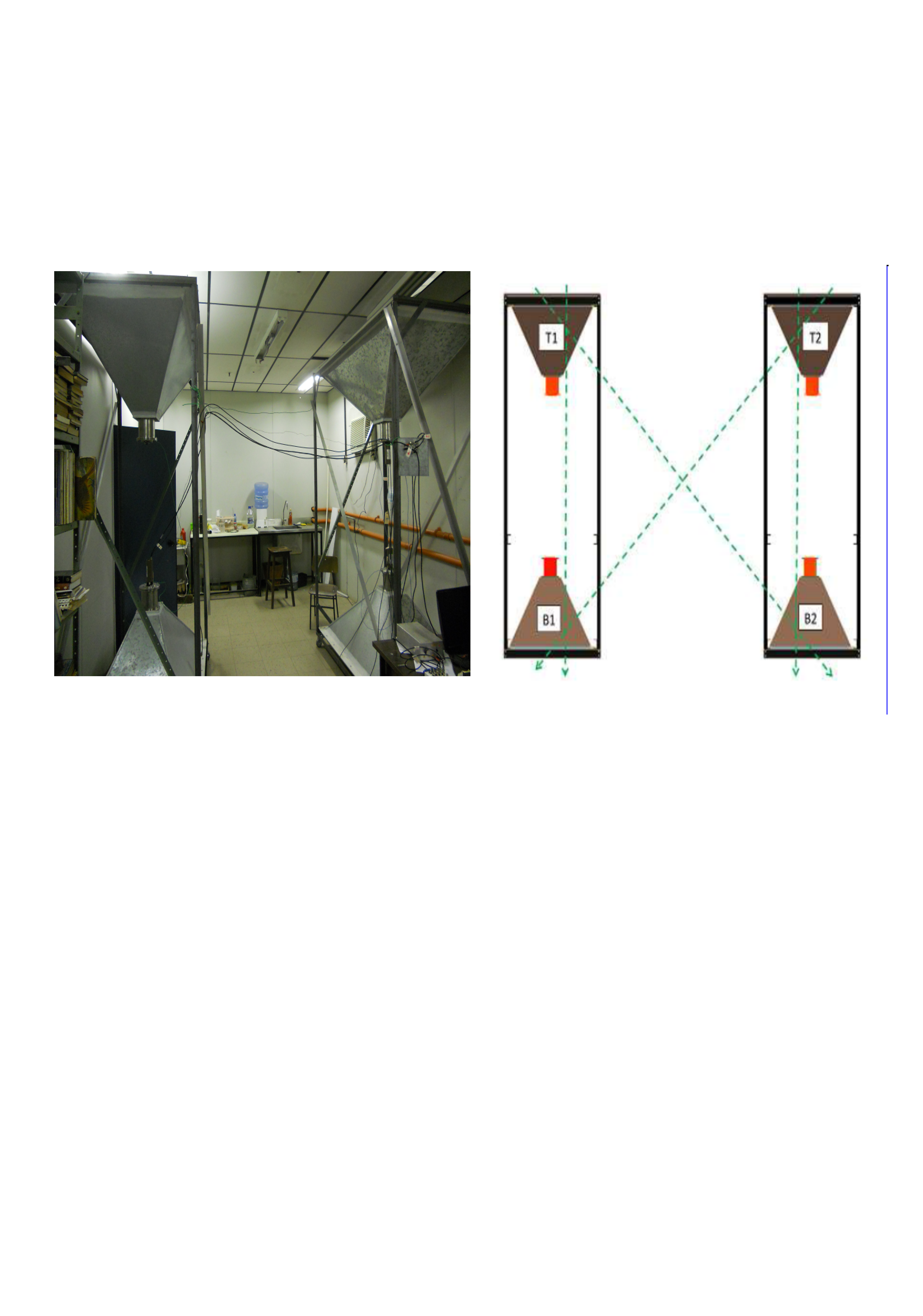}
\vspace*{-8.5cm}
\caption{Left: Photograph of the New-Tupi detector, located in the building of the Physical Institute of the Universudade Federal Fluminense, Niteroi, Rio de Janeiro, Brazil.
 Right: Scheme of the New-Tupi detector. The configuration in telescope mode (coincidence rates between two detectors) allowed to obtain the muon flux
 from three directions, the vertical, west and east, the last two with a inclination of 45 degrees, as is indicated by the dash lines. In addition, there is the scaler mode, where the single hit rates of all of the four PMTs are recorded at a fixed rate.
}
\label{fig8}
\end{figure}

Fig. 1 (left) shows a photography of the New-Tupi detector. It can operate
in  the telescope mode, for this purpose,
 the four scintillators are placed in pairs, with T1 (top) and B1 (bottom) and T2
and B2 respectively, as shown in Fig. 1 (right), this layout allow to obtain the
muon flux from three directions, the vertical, west and east, the last two
with an inclination around 45 degrees. The telescopes registers the coincidence
rate among T1 and B1; T2 and B2 (both are vertical incidence); and the cross
coincidences T1 and B2 (west incidence) and the T2 and B1 (east incidence).
Both vertical and lateral separation between the detectors are 2.83 m.

In parallel with the telescope mode, the New-Tupi detector can also operate in scaler mode or mode called also as single particle technique \cite{obrian76,morello84,aglietta96}, where the single hit rates of all of the four PMTs were recorded once a second.
This mode allows to search by particle excess, at fixed time intervals, even when the source is out of the field of view of the telescopes, because the field of view of a single detector is bigger than of field of view of a telescope. However, the efficiency of particle detection in scaler mode decreases as the zenith angle of the incident particle increases, due to atmospheric absorption. Thus the scaler method is limited to incident particles with a zenith angle not more than 60 degrees.

\acknowledgments

We acknowledge the Space Weather Prediction Center (SWPC);
the CACTus COR2 CME list and the NMDB database, founded under
the European Union’s FP7 programme (contract no. 213007) for
providing data. The data of the South Pole NM; McMurdo NM and Thule NM
are provided by the University of Delaware with support from the U.S. National Science Foundation under grant ANT-0838839.

This work is supported by the National Council for Research (CNPq) of
Brazil, under Grant 306605/2009-0 and Fundacao de Amparo a Pesquisa do Estado do Rio de Janeiro (FAPERJ), under Grant 08458.009577/2011-81 and E-26/101.649/2011.


\newpage

\begin{references}

\bibitem{simpson00}J. A. Simpson, J. A.  Space Science Reviews, 93, 11 (2000)
\bibitem{oh12}Oh et al., Space Weather, 10, S05004 (2012).
\bibitem{firoz10}K. A. Firoz et al., Journal of Geophysics Research, 115, A09105 (2010)
\bibitem{gopal12} N. Gopalswamy, et al., arXiv:1205.0688 [astro-ph.SR]
\bibitem{thakur14}N. Thakur et al., arXiv:1406.7172 [astro-ph.SR]
\bibitem{andriopoulou11}M. Andriopoulou et al., Solar Physics, 269, 155 (2011).
\bibitem{bartol1}http://neutronm.bartol.udel.edu/
\bibitem{bartol2}http://www.bartol.udel.edu/~takao/icetop/01m/gle73.html
\bibitem{robb04}E. Robbrecht and D. Berghmans, A$\&$A, 425, 1097 (2004) 
\bibitem{wilcox65}J. M. Wilcox, N. F. Ness,  Journal of Geophysical Research, 70, 5793 (1965)
\bibitem{svestka68}Z. \~{S}vestka, Solar Physics, 4, 361 (1968)
\bibitem{olbert54}S. Olbert, PhRv, 96, 1400 (1054).
\bibitem{reyes05}R. De Los Reyes et at. et al., Int. J. of Mod. Phys. A, 20, 7006 (2005).
\bibitem{augusto15}C. R. A. Augusto et al, Astrophys. J., 805, 89 (2015).
\bibitem{noaa}$http://www.ngdc.noaa.gov/geomag-web/\# igrfwmm$
\bibitem{israelevich01}P. L. Israelevich, A. I. Ershkovich, and N. A. Tsyganenko, J. Geophys. Res., 106, 25919 (2001)
\bibitem{foullon09}C. Foullon, et al., Solar Physics, 259, 389 (2009)
\bibitem{mursula03}K. Mursula and T. Hiltula, Geophys. Res. Lett., 30, 2135 (2003)
\bibitem{kuwabara06}T. Kuwabara, et al. SPACE WEATHER, 4, S10001 (2006).
\bibitem{mavro12}H. Mavromichalaki, Advances in Solar and Solar-Terrestrial Physics,2012: 135-161 ISBN: 978-81-308-0483-5
Editors: Georgeta Maris and Crisan Demetrescu. 
\bibitem{villante79}U. Villante, R. Bruno, F. mariani, L.F. Burlaga and N.F. Ness,
J. Geophys. Res, 84, 6641 (1979).



\bibitem{usoskin}I. Usoskin,\textit{Oulu neutron monitor cosmic ray data}, 
Volume 89 de Sodankylä Geophysical Observatory publications, Sodankylä Geophysical Observatory publications, Oulu University Press, 2001

\bibitem{oulu}http://cosmicrays.oulu.fi/GLE.html

\bibitem{cliver95}E. W. Cliver, Transactions American Geophysical Union 76: 
doi:10.1029/95EO00035

\bibitem{cane96}H. V. Cane: 1996, in: Ramaty, R., Mandzhavidze, N., and Hua, X. M. (eds.), High Energy Solar Physics, AIP, Woodberry, N.Y., pp. 124–130.

\bibitem{augusto15b}C.R. Augusto et al., Publ Astron Soc Jpn, published online November 19, 2015  doi: 10.1093/pasj/psv111

\bibitem{augusto15c}C. R. Augusto et al., arXiv:1507.03954 [astro-ph.SR]
\bibitem{barton97}C. E. Barton,  J. Geomagn. Geoelectr. 4942002, 123 (1997).

\bibitem{obrian76}S. O'Brian and N. A. Porter, Astrophys. Space Sci. 42, 73 (1976)
\bibitem{morello84}C. Morello, G. Navarra, and L. Periale, Nuovo Cimento, 7C, 682 (1984)
\bibitem{aglietta96}M. Aglietta et al. , Astrophys. J., 469, 305 (1996)




\end{references}
\end{document}